\documentclass[aps,superscriptaddress,showpacs,prd,nofootinbib,eqsecnum]{revtex4-1}

\usepackage{graphics} 
  
\usepackage{xcolor}
\usepackage{epsfig} 
\input{epsf}

\usepackage{natbib} 
\usepackage[colorlinks=true,citecolor=blue,linkcolor=blue,urlcolor=blue]{
hyperref}
\setcitestyle{square, numbers, comma, sort&compress}

\def\be{\begin{equation}}
\def\ee{\end{equation}}
\def\bea{\begin{eqnarray}}
\def\eea{\end{eqnarray}}

\begin{document} 
\title {  Three dimensional Yukawa models and CFTs  at strong and weak couplings}

\author{Marcus Benghi Pinto} \email{marcus.benghi@ufsc.br}
\affiliation{Departamento de F\'{\i}sica, Universidade Federal de Santa
  Catarina, 88040-900 Florian\'{o}polis, SC, Brazil}

 \begin{abstract} 
 The massless three dimensional Gross-Neveu-Yukawa (GNY)  and Nambu--Jona-Lasinio--Yukawa (NJLY)  models at finite temperatures are analyzed within the mean field framework considering all coupling values. When the number of Dirac fermions is taken to be $N_f=1/4$ (GNY) and $N_f=1/2$ (NJLY) these models relate to  
 the supersymmetric Wess-Zumino (WZ) theory with  cubic superpotential and one superfield. In this case the results show that the strong-weak entropy density ratio  decreases from the Stefan-Boltzmann value, in the weak limit,  to $s/s_{free}=31/35$ at strong couplings. This value agrees with the one recently  obtained  by applying the large-$N$ approximation to the  supersymmetric  $O(N)$ WZ  model with  quartic superpotential and $N$ superfields. When $N_f=0$ one obtains $s/s_{free}=4/5$ recovering, as expected, the ratio predicted in the context of the   $O(N)$ scalar model. However, contrary to the   $O(N)$ WZ model the simple Yukawa models analyzed here do not behave as CFTs for all couplings since the conformal measure exactly vanishes only at the extreme weak and strong limits although the speed of sound indicates that the  deviation, at intermediate couplings, appears to be rather small.  By comparing the thermal masses behavior in each case one can trace this difference as being a consequence that in the GNY/NJLY case the fermionic mass vanishes for all couplings  while within the  $O(N)$ WZ it only vanishes at the weak and strong limits. On the other hand,  the Yukawa bosonic  dimensionless masses  display a more universal   behavior decreasing from $2 \ln [(1+\sqrt{5})/2]$, at infinite coupling, to zero (at vanishing coupling).  
 \end{abstract} 
 
 
\maketitle
 
\section{Introduction} 
Superrenormalizable massless three dimensional theories at finite temperatures provide a useful framework to analyze CFT candidates at all coupling values owing to the fact that any dimensionful coupling can be expressed in terms of the temperature so that strong coupling values can be generated by considering low temperatures while weak coupling values are obtained at high temperatures.  Recently this interesting feature has been explored in  the context of the scalar  $O(N)$ model whose dimensionful coupling, $\lambda$,  has been combined with the temperature in  the dimensionless ratio $\lambda/T$ allowing for  investigations at all couplings (including  infinite values as  $T \to 0$) \cite {Romatschke:2019ybu}. That application, which has been carried out at the large-$N$ limit, has produced some interesting results such as predicting that the value of the entropy density  decreases from the Stefan-Boltzmann value at $\lambda=0$ to exactly 4/5 of the Stefan-Boltzmann limit at $\lambda=\infty$. 
Recalling that  in the gauge/gravity  duality  context   the  result  for  the  entropy  density  of  strongly coupled ${\cal N}= 4$  SYM in four dimensions, and  at large-$N$, is exactly 3/4 of the Stefan-Boltzmann limit \cite {Gubser:1998nz} one may argue that these two theories share a similar strong-weak relation as far as the entropy ratio is concerned. However, as noted in Ref. \cite {Romatschke:2019ybu}, it is important to mention that the scalar  $O(N)$ results were obtained just by applying the standard thermo field machinery to a rather simple model without any invocation of gauge/gravity duality. Regarding the results obtained 
 in the  scalar case \cite {Romatschke:2019ybu}
 it becomes  natural to ask how the consideration of fermionic degrees of freedom would eventually affect the 4/5 entropy density ratio obtained with such  purely bosonic theory. To  answer this question DeWolfe and Romatschke \cite {DeWolfe:2019etx} have extended the scalar  $O(N)$ application to the three dimensional supersymmetric  $O(N)$ Wess-Zumino model \cite {Wess:1974tw}, which displays a quartic Yukawa vertex, at large-$N$. One of the main outcomes of this study is that, at infinite coupling, the strong-weak ratio is  exactly 31/35 when an equal number of fermions ($F$)  and bosons ($B$) is considered. At the same time by taking the extremum case where $F \to 0$ (or $B \to \infty$) the value $s/s_{free}=4/5$ is recovered while $s/s_{free}=1$ is obtained when  $B \to 0$ (or $F \to \infty$) so that the depending on the balance between fermions and bosons the entropy density ratio is bounded to lie between 4/5 and 1. The aim of the present work is to investigate how three dimensional massless theories, with a three linear Yukawa vertex, behave at all coupling values by comparing the results with the ones obtained in the  $O(N)$ WZ case as well as to identify the physical source of possible differences. With this purpose the thermodynamics of the  three dimensional massless  Gross-Neveu-Yukawa (GNY) \cite {Hasenfratz:1991it,ZinnJustin:1991yn} and  Nambu--Jona-Lasinio--Yukawa (NJLY) \cite {Hasenfratz:1991it} models  will be considered at the mean field (one loop) level in order to evaluate thermodynamical quantities such as the strong-weak entropy density ratio, the conformal measure as well as the speed of sound squared.  It goes without saying that the interaction between fermions and bosons via a trilinear Yukawa vertex is of utmost importance to describe a plethora of physical situations such as the ones covered by the standard model of elementary particles, the Walecka model for nuclear matter  \cite {Walecka:1974qa,*Serot:1984ey} and the quark-meson model \cite {GellMann:1960np} among many other examples in different areas of Physics. At the same time the YGN and YNJL models are related to the four-fermion Gross-Neveu (GN) \cite {Gross:1974jv} and Nambu--Jona-Lasinio (NJL) \cite {Nambu:1961tp,*Nambu:1961fr} theories which are often used as model approximations to quantum chromodynamics (QCD) in studies related to the chiral transition. 
 With respect to supersymmetric models it is also important to remark that the YGN/YNJL theories considered here relate to the  WZ model with  one superfield and a cubic superpotential \cite {Fei:2016sgs} while the $O(N)$ WZ version considered in Ref. \cite {DeWolfe:2019etx} describes $N$ superfields interacting via a quartic superpotential.
 The work is organized as follows. In the next section the YGN and YNJL models are presented and their free energy densities are evaluated within the mean field approximation (MFA).   The pressure and other relevant thermodynamical quantities are defined together with the gap equations in Sec. III. Analytical and numerical results at all coupling values are presented and discussed in Sec. IV. Finally, Sec. V contains the conclusions and perspectives.

\section{The Yukawa models}

To facilitate further comparisons let us  first recall that the version of the  WZ model  analyzed in Ref. \cite {DeWolfe:2019etx} describes $N$ superfields whose dynamics is dictated by a quartic superpotential. Such a a theory can be described by the lagrangian density
\begin{equation}
{\cal L}_{WZ}= \frac{1}{2} (\partial_\mu \phi_a)(\partial_\mu \phi_a) + \frac{1}{2} {\overline \psi}_a (i \not \! \partial) \psi_a -\frac{2\lambda}{N}\phi_b \phi_b {\overline \psi}_a \psi_b - \frac{8 \lambda^2}{N} (\phi_a \phi_a)^3\;,
\label{WZ4}
\end{equation}
where $\phi_a$ ($a=1,...,N$) represents $N$-component real scalars while $\psi_a$ represents a $N$-component Majorana spinor in $2+1d$. Note that  the original dimensionless couplings have already been rescaled by $1/N$ in order to allow  for the implementation of large-$N$ evaluations. Also, for future reference, remark that bosons self interact through a sextic vertex while bosons and fermions interact through a quartic Yukawa vertex. At large-$N$ the conformal measure for such a theory vanishes for all values of $\lambda$ so that one may say that the model describes  a ``pure"  CFT just like the scalar $O(N)$ model with a sextic vertex analyzed in Ref. \cite {Romatschke:2019ybu} (see Ref. \cite {DeWolfe:2019etx} for further details).

\subsection{The Yukawa-Gross-Neveu model}
In Minkowski space the massless Yukawa-Gross-Neveu model\footnote {Sometimes called Higgs-Yukawa model.} describing  one scalar, $\phi$, and $N_f$ four component Dirac fermions, $\psi_f$ ($f=1,...,N_f$), can be described by the lagrangian density \cite {Hasenfratz:1991it,ZinnJustin:1991yn}
\begin{equation}
{\cal L}_{YGN} = \frac{1}{2} (\partial_\mu \phi)^2 + {\overline \psi}_f (i \not \! \partial) \psi_f - g_1 \phi {\overline \psi}_f \psi_f - \frac{g_2}{8} \phi^4 \;,
\end{equation}
which is invariant under the discrete transformations\footnote {Note that in $2+1d$ this is true only one considers $4\times 4$ Dirac matrices as we do here. See Refs \cite {Appelquist:1986fd, Rosenstein:1990nm} for details.  } $\psi_f \to \gamma_5 \psi_f$ and $\phi \to -\phi$.
Note that in $2+1$ dimensions the couplings have canonical dimensions $[g_1]=1/2$ and $[g_2]=1$  and the theory is  superrenormalizable. Also, since there are no logarithmic divergencies the $\beta$ functions vanish which is a further requirement for CFTs. For our purposes the large-$N$ approximation does not seem to be the most appropriate tool to treat this model not only because there is just one boson but also because we shall relate its results to the WZ model with only one superfield as will be further discussed. In this case one may alternatively consider the MFA which, by considering only one loop (direct) contributions, relates not only to the large-$N$  itself but also to the traditional Hartree approximation.
One may implement the MFA by defining a space-time independent classical field, $\sigma_c = \langle \phi^2 \rangle_0$,  while considering the mean field approximation $\phi^4 \simeq 2 \sigma_c \phi^2 - \sigma_c^2$. One can then shift $\phi \to \phi^\prime +\phi_c$ and reexpand neglecting all terms linear in $\phi^\prime$ since they either produce non 1PI contributions or 1PI terms which only contribute beyond the (one loop) mean field level.  After doing that, dropping the superscript in $\phi^\prime$,   rescaling $\sigma_c \to  g_2 \sigma_c/2$, and $\phi_c \to g_1 \phi_c$  the lagrangian density within the MFA can be written as 

\begin{equation}
{\cal L}_{YGN} \simeq \frac{1}{2} \left [ (\partial_\mu \phi)^2 - \sigma_c \phi^2 \right ]+ {\overline \psi}_f\left [ (i \not \! \partial)  - \phi_c \right ]  \psi_f +
\frac{\sigma_c^2}{2 g_2}  - \frac{\sigma_c \phi_c^2}{2 g_1^2}\; .
\end{equation}
Since now the bosonic and fermionic integrals are gaussian   the free energy density can be easily evaluate by standard methods \cite {Bailin} yielding
\begin{equation}
{\cal F}_{YGN} (\sigma_c, \phi_c)= - \frac{\sigma_c^2}{2 g_2} + \frac{\sigma_c \phi_c^2}{2 g_1^2}  - \frac{i}{2} \int \frac{d^3 p}{(2\pi)^3} \ln (p^2 - \sigma_c)+ 2 N_f i \int \frac{d^3 p}{(2\pi)^3} \ln (p^2 - \phi_c^2)\;.
\end{equation}
To perform finite temperature evaluations in the Matsubara imaginary time formalism \cite {Bailin, Laine:2016hma, Kapusta:2006pm} one needs to rewrite the zeroth momentum component as  $p_0 \to i \omega_n$ where $\omega_n$ represents the Matsubara's frequencies which are defined as  $\omega_{F,n} = (2n+1)\pi T$, for fermions, and $\omega_{B,n} = 2\pi Tn$ for bosons where $n=0, \pm 1, \pm 2,...$.
Also, in order to sum over the Matsubara's frequencies the   integrals over loops need to be modified according to \cite {Laine:2016hma, Kapusta:2006pm}
\begin{equation}
\int \frac {d^3 p} {(2\pi)^3} \to i T \,
\hbox{$\sum$}\!\!\!\!\!\!\!\int_{\bf p} \equiv i T\left ( \frac {e^{\gamma_E} M^2}{4\pi}
\right)^{\epsilon} \sum_{n=-\infty}^{+\infty} \int
\frac {d^{2-2 \epsilon} {\bf p}}{(2\pi)^2} \;,
\end{equation}
 where $\gamma_E$ is the Euler-Mascheroni constant and $M$
is the $\overline {\rm MS}$ arbitrary regularization energy scale. One then gets
\begin{equation}
{\cal F}_{YGN} (\sigma_c, \phi_c)= - \frac{\sigma_c^2}{2 g_2} + \frac{\sigma_c \phi_c^2}{2 g_1^2} + \frac{1}{2} J_B(\sqrt{\sigma_c}) - 2 N_f J_F(\phi_c)\;,
\label {freeYGN}
\end{equation}
where $J_B(\sigma_c)$  and $J_F(\phi_c)$ represent thermal integrals. In terms of the dispersions $\omega_B^2= {\bf p}^2 + \sigma_c$ and $\omega_F^2= {\bf p}^2 + \phi_c^2$ these integrals read
\begin{equation}
J_B(\sqrt{\sigma_c}) = \hbox{$\sum$}\!\!\!\!\!\!\!\int_{\bf p} \ln [\omega_{B,n}^2 + \omega_B^2]\;\;\;\;\;\;\;{\rm and}\;\;\;\;\;\;\;J_F(\phi_c) = \hbox{$\sum$}\!\!\!\!\!\!\!\int_{\bf p} \ln [\omega_{F,n}^2 + \omega^2_F] \;.
\end{equation}

In 2+1 dimensions both integrals, which are finite and scale independent within dimensional regularization, can be expressed in a compact form in terms of polylogarithmic functions as 
\begin{equation}
J_B(\sigma_c) = - \frac{\sigma_c^{3/2}}{6\pi} - \sqrt{\sigma_c} \frac{T^2}{\pi} {\rm Li}_2 [ e^{-\sqrt{\sigma_c}/T}] - \frac {T^3}{\pi} {\rm Li}_3 [ e^{-\sqrt{\sigma_c}/T}]\;,
\label{JB}
\end{equation}
and 
\begin{equation}
J_F(\phi_c) = - \frac{\phi_c^3}{6\pi} - \phi_c \frac{T^2}{\pi} {\rm Li}_2 [ - e^{-\phi_c/T}] - \frac {T^3}{\pi} {\rm Li}_3 [- e^{-\phi_c/T}] \;.
\label{JF}
\end{equation}
Regarding the relation between supersymmetric models and the YGN model it is interesting to observe that the balance between bosons and fermions is dictated by the coefficients of $J_B$ and $J_F$ appearing in Eq. (\ref {freeYGN}). One immediately notices that, in particular, the value $N_f=1/4$ represents the relevant case as far as comparisons with the WZ results of Ref. \cite {DeWolfe:2019etx} are concerned. Indeed, as suggested in Ref. \cite {Fei:2016sgs} one may define $N=4 N_f$ so that $N$ is the number of two component Majorana fermions in 2+1 dimensions. In this vein it is worth to recall the suggestion that a minimal ${\cal N}=1$ SCFT containing a single two-component Majorana fermion may exist in $2+1d$. To describe such a theory the following Lagrangian density, in Minkowski space, has been proposed \cite {Iliesiu:2015qra, Grover:2013rc, Bashkirov:2013vya}
\begin{equation}
{\cal L}_{{\cal N}=1}= \frac{1}{2} (\partial_\mu \phi)^2 + \frac{1}{2} {\overline \psi} (i \not \! \partial) \psi -\frac{\lambda}{2}\phi {\overline \psi} \psi - \frac{\lambda^2}{8} \phi^4\;.
\label{WZ3}
\end{equation}
When the coupling relation $\lambda^2=g_2=g_1^2$ is satisfied this  ${\cal N}=1$ Wess-Zumino model belongs to the same universality class as the YGN at $N_f=1/4$ \cite {Fei:2016sgs,Sonoda:2011qd}.  

\subsection{The Yukawa--Nambu--Jona-Lasinio model}
The massless Yukawa--Nambu--Jona-Lasinio  lagrangian density describing  two scalars, $\phi_i$ (i=1,2), and $N_f$ four component Dirac fermions, $\psi_f$ ($f=1,...,N_f$),  can be written as \cite {Hasenfratz:1991it}

\begin{equation}
{\cal L}_{YNJL} = \frac{1}{2} (\partial_\mu \Phi^*)(\partial_\mu \Phi) + {\overline \psi}_f \left[(i \not \! \partial)  - g_1 (\phi_1 +i \gamma_5 \phi_2)\right ]\psi_f - \frac{g_2}{8} (\Phi^* \Phi)^2\; ,
\end{equation}
where $\Phi = \phi_1 + i \phi_2$ such that the theory is   invariant under the continuous transformations $\psi_f \to e^{i\alpha \gamma_5} \psi_f$ and $\Phi \to e^{i2\alpha}\Phi$. As in the YGN case the interactions can be linearized by using the MFA
\begin{equation}
(\Phi^* \Phi)^2 \simeq 2 \Phi^*\Phi(\xi_c^*\xi_c)^{1/2} - \xi_c^*\xi_c \;,
\end{equation}
where $\xi_c = \sigma_c + i\pi_c$ with  $\sigma_c = \langle \phi_1^2 \rangle_0$ and $\pi_c = \langle \phi_2^2 \rangle_0$. Then, shifting $\phi_i \to \phi_i^\prime  +\phi_{i,c}$ ($i=1,2$) and proceeding as in the YGN case one obtains the MFA free energy density
\begin{eqnarray}
{\cal F}_{YNJL}(\sigma_c,\pi_c,\phi_{1,c},\phi_{2,c})& =& - \frac{g_2}{8} (\sigma_c^2 +\pi_c^2) + \frac{g_2}{4}(\sigma_c^2 +\pi_c^2)^{1/2}(\phi_{1,c}^2+\phi_{2,c}^2) -
i \int \frac{d^3 p}{(2\pi)^3} \ln \left [ p^2 - \frac{g_2}{2}(\sigma_c^2 +\pi_c^2)^{1/2} \right ] \nonumber \\
&+& 2 N_f i \int \frac{d^3 p}{(2\pi)^3} \ln[p^2 - g_1^2 (\phi_{1,c}^2+\phi_{2,c}^2)]\;.
\end{eqnarray}
Due to the apparent symmetry the free energy density can be more conveniently examined at the particular  points $\pi_c=0$ and $\phi_{2,c}=0$. Next, one can define $\phi_{1,c}=\phi_c$ while rescaling $\phi_c \to \phi_c/g_1$ and $\sigma_c \to (2/g_2) \sigma_c$  to finally write 
\begin{equation}
{\cal F}_{YNJL} (\sigma_c, \phi_c)= - \frac{\sigma_c^2}{2 g_2} + \frac{\sigma_c \phi_c^2}{2 g_1^2} +  J_B(\sqrt{\sigma_c}) - 2 N_f J_F(\phi_c)\;,
\end{equation}
where $J_B(\sqrt{\sigma_c})$ and $J_F(\phi_c)$ are given by Eqs. (\ref {JB}) and (\ref {JF}). Regarding analogous  SUSY models note that we now have two bosons and the definition $N=4 N_f=2$ sets $N_f=1/2$ as the relevant value when relating the YNJL to  the Wess-Zumino ${\cal N}=2$ theory of a chiral superfield with cubic superpotential  (see Ref. \cite {Fei:2016sgs} and references therein for more details).

\section{Thermodynamics}
In order to easily explore the thermodynamics of the YGN and YNJL models let us  rewrite the free energy density in terms of the number of bosons, $N_b$, as 
\begin{equation}
{\cal F} (\sigma_c, \phi_c)= - \frac{\sigma_c^2}{2 g_2} + \frac{\sigma_c \phi_c^2}{2 g_1^2} +  \frac{N_b}{2}J_B(\sqrt{\sigma_c}) - 2 N_f J_F(\phi_c)\;,
\end{equation}
which is a form appropriate to treat both situations by selecting $N_b=1$ and $N_f=1/4$ (YGN) or $N_b=2$ and $N_f=1/2$  (YNJL).
The pressure can be obtained from the relation $P = - {\cal F}({\overline \sigma}, {\overline \phi})$ where ${\overline \sigma}$ and  ${\overline \phi}$ satisfy the ``gap" equations $ \partial {\cal F}/\partial \sigma_c =0$ and $\partial {\cal F}/\partial \phi_c=0$. One then obtains the coupled equations
\begin{equation}
\frac{{\overline \phi}^2}{2 g_1^2} = \frac{\overline \sigma}{g_2} +\frac{N_b}{8\pi} \left [ {\sqrt{\overline \sigma}} +2 T \ln \left ( 1 - e^{-\sqrt{\overline \sigma}/T} \right )\right ]\;,
\label{gap1}
\end{equation}
and 
\begin{equation}
{\overline \sigma}\frac{\overline \phi}{g_1^2} = - N_f \frac {\overline \phi}{\pi} \left [ {\overline \phi} + 2 T \ln \left ( 1 + e^{-{\overline \phi}/T} \right) \right ]\;,
\label{gap2}
\end{equation}
where the last equation  has not been  simplified since the trivial solution ${\overline \phi}=0$ will prove to be useful in the sequel.
Next, let us write the entropy density $s= \partial P /\partial T$ as a sum of the bosonic and fermionic contributions $ s(T)= s_B(T) +  s_F(T)$ where
\begin{equation}
s_B(T)= \frac{N_b}{2\pi} \left \{ 3 \sqrt{\overline \sigma}  T  {\rm Li}_2[e^{-\sqrt{\overline \sigma}/T}] + 3 T^2  {\rm Li}_3 [e^{-\sqrt{\overline \sigma}/T}] - {\overline \sigma}\ln [ 1- e^{-\sqrt{\overline \sigma}/T}] \right \}\;,
\end{equation}
and 
\begin{equation}
s_F(T)= - N_f\frac{2}{\pi} \left \{ 3 {\overline \phi}  T  {\rm Li}_2[-e^{-{\overline \phi}/T}] + 3  T^2  {\rm Li}_3 [-e^{-{\overline \phi}/T}] - 
{\overline \phi}^2\ln [ 1+ e^{-{\overline \phi}/T}] \right \}\;.
\end{equation}
The above equations are guaranteed to be thermodynamically consistent thanks to the gap equations which 
eliminate the crossed terms $ (\partial {\overline \phi}/\partial T)
(\partial P/\partial {\overline \phi})$ and $ (\partial {\overline \sigma}/\partial T)
(\partial P/\partial {\overline \sigma})$.
The Stefan-Boltzmann limit can be  easily obtained by taking $\overline \sigma=0$ and $\overline \phi =0$ so that we can write $s_{free} = s_{B,free} + s_{F,free}$ where
\begin{equation}
s_{B,free} = T^2 N_b \frac{3 \zeta(3)}{2 \pi}\;\;\;\;\;\;{\rm and}\;\;\;\;\;\;s_{F,free} = T^2 N_f \frac{9 \zeta(3)}{2 \pi} \;,
\end{equation}
implying that $s_{F,free} =  3 (N_f /N_b) s_{B,free}$. At the same time these relations allow us to trivially set  $P_{free} = T s_{free}/3$. Using these results one can easily obtain the energy density ${\cal E}= -P + s T$, the trace anomaly $\Delta = ({\cal E} - 2P)$ as well as the conformal measure
\begin{equation}
{\cal C} = \frac{\Delta}{\cal E} \;,
\end{equation}
and the  speed of sound squared
\begin{equation}
V_s^2 = \frac {\partial P}{\partial {\cal E}} = \frac{s}{C_v}\;,
\end{equation}
 where $C_v= T(\partial s)/(\partial T)$ is the specific heat.

\section{Results}
To examine  thermodynamical quantities we first need to solve the gap equations (\ref {gap1}) and (\ref {gap2}). To do that let us start by defining the dimensionless thermal masses $ m_F = {\overline \phi}/T$ and $m_B =\sqrt {\overline \sigma}/T$ so that the gap equations become
\begin{equation}
\frac{T}{2 g_1^2} m_F^2 = \frac{T}{g_2} m_B^2 + \frac{N_b}{8 \pi}\left \{ m_B  + 2 \ln[ 1- e^{-m_B}] \right \}\;,
\label{massaB}
\end{equation}
and 
\begin{equation}
\frac{T}{g_1^2} m_B^2 m_F = - m_F \frac{N_f}{\pi} \left \{ m_F + 2 \ln[ 1+ e^{-m_F}] \right \}\;,
\label{massaF}
\end{equation}
where we again have not cancelled an overall factor of $m_F$ in the last equation. 

We can now solve the above equations at the strong ($T\to 0$) and weak ($T \to \infty$) coupling limits starting with the former. In this case the gap equations decouple and the first one sets
\begin{equation}
m_B = 2 \ln \Phi \;,
\end{equation}
where $\Phi$ represents the golden ratio $(1+\sqrt{5})/2$ exactly as in the  $O(N)$ scalar case  \cite{Romatschke:2019ybu}. The second equation has two solutions: the first one is the trivial $m_F=0$ while the second gives the complex $m_F = \pm (2 \pi i)/3$ which we discard. As expected, at the Stefan-Boltzmann limit the solutions are $m_F = m_B \equiv 0$. As it was numerically checked $m_F =0$ for all temperatures so that one ends up with only the following equation for $m_B$
\begin{equation}
 \frac{T}{g_2} m_B^2 = - \frac{N_b}{8 \pi}\left \{ m_B  + 2 \ln[ 1- e^{-m_B}] \right \}\;.
 \label{mB}
\end{equation}

\begin{figure}[htb]
 \centerline{ \epsfig{file=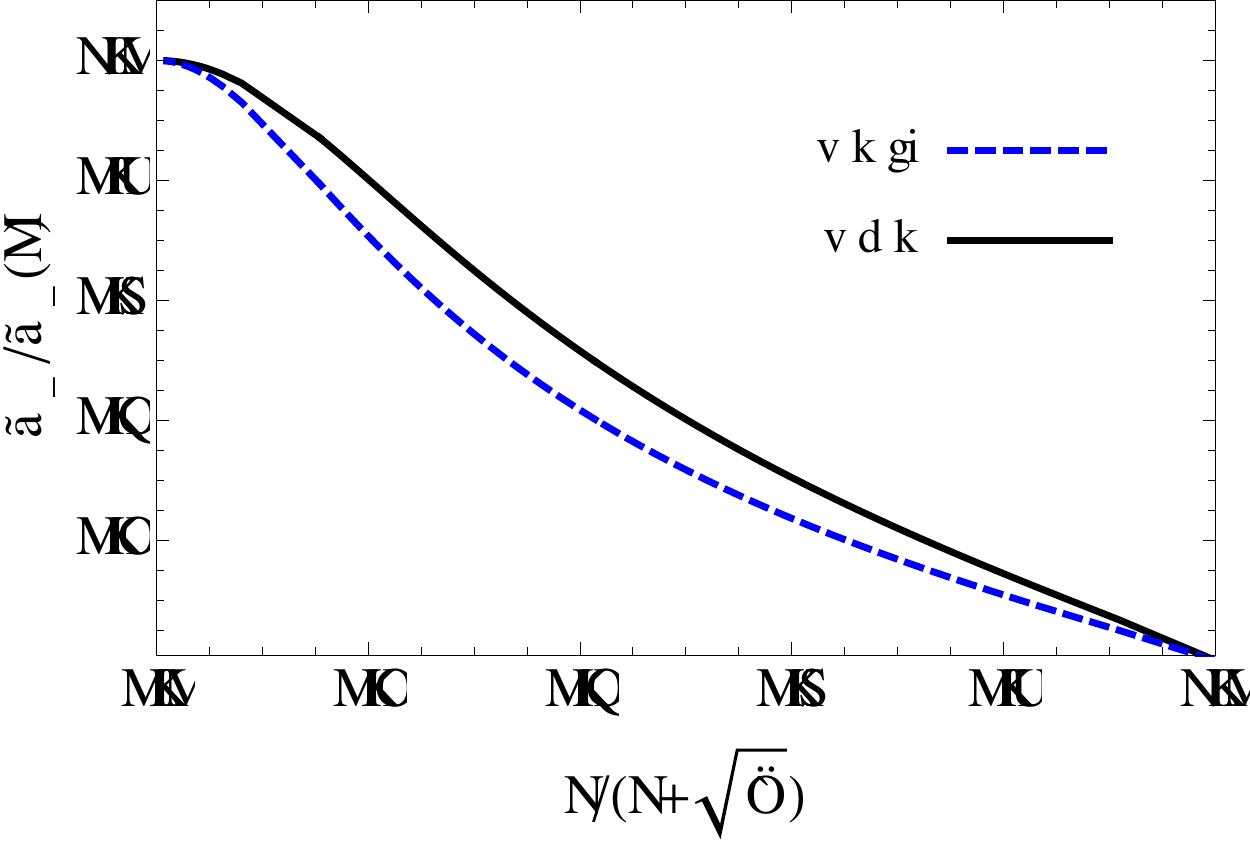,width=0.45\linewidth,angle=0}}
\caption{The YGN and YNJL bosonic masses, normalized by $m_B(0)=2 \ln \Phi$, as a function of the quantity $1/(1+\sqrt{{\hat g}})$ which  ranges from 0 (strong coupling or low $T$ limit) to 1  (weak coupling or high $T$ limit). The masses vary from $m_B=2 \ln \Phi \simeq 0.962$ at ${\hat g} = g_2/T  \to \infty$ to $m_B=0$ at the Stefan-Boltzmann limit,  ${\hat g} = g_2/T  \to 0$.}
\label{Fig1}
\end{figure}

 Defining the dimensionless coupling ${\hat g} = g_2/T$ one can then investigate $m_B$ in between the strong (${\hat g} \to \infty$) and weak (${\hat g} \to 0$) limits by plotting the bosonic mass in the compactified interval $(1+\sqrt{{\hat g}})^{-1} \in [0,1]$ as in Fig. \ref{Fig1}.
Then, taking into account that $m_F=0$ for all couplings the pressure  can be written in a more compact form as 
\begin{equation} 
P= T^3 \left \{ \frac{m_B^4}{ 2 {\hat g}} +\frac{N_b}{2\pi} \left [\frac {m_B^3}{6} + m_B {\rm Li}_2[e^{-m_B}] +  {\rm Li}_3[e^{-m_B}]\right ] \right \} +
N_f T^3 \frac{3 \zeta(3)}{2\pi}\;,
\end{equation}
where the last term is just $P_{F,free}$. Then, the  entropy density reads
\begin{equation}
s(T) =  N_b \frac{T^2}{2\pi} \left \{ 3 m_B {\rm Li}_2[e^{-m_B}] + 3  {\rm Li}_3 [e^{-m_B}] - m_B^2 \ln [ 1- e^{-m_B}] \right \} + N_f T^2 \frac{9 \zeta(3)}{2\pi}\;,
\end{equation}
where the last term represents $s_{F,free}$.

\begin{figure}[htb]
\centerline{ \epsfig{file=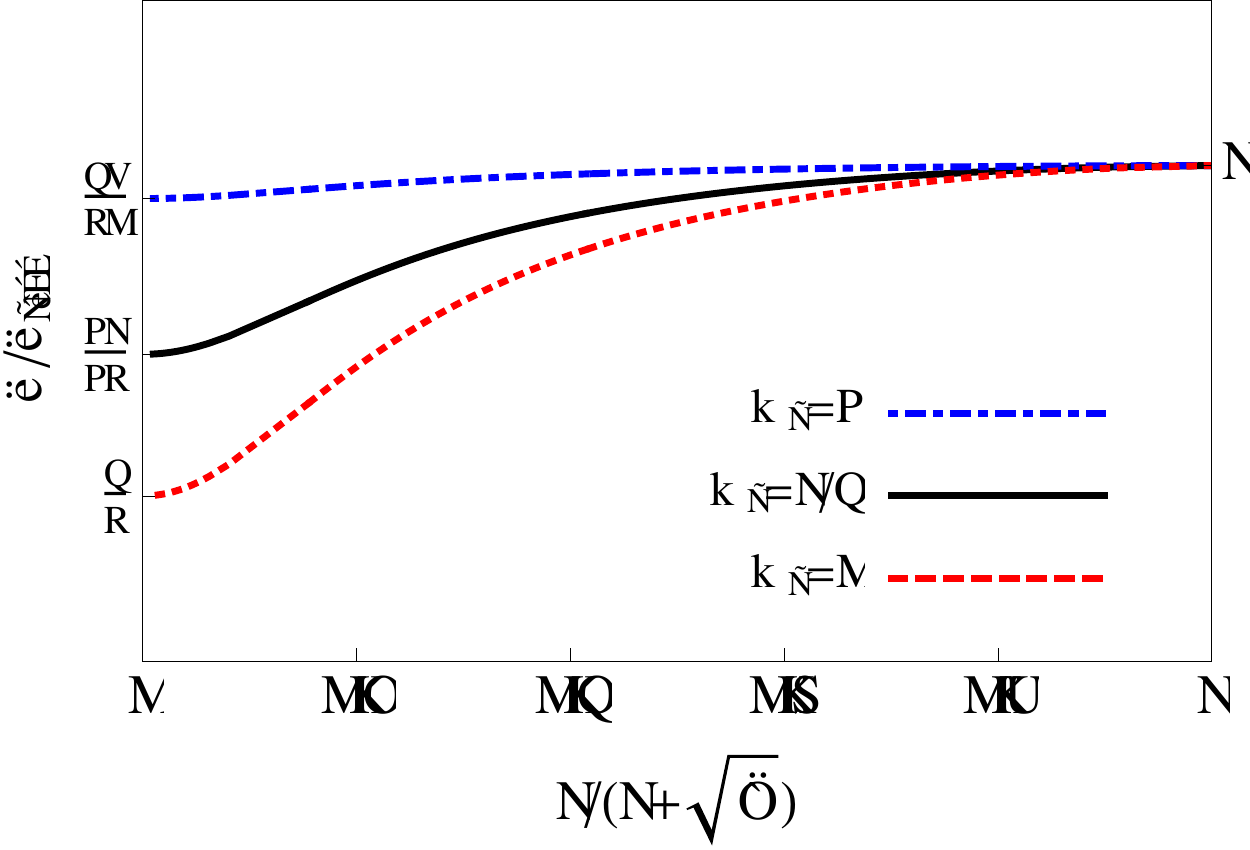,width=0.45\linewidth,angle=0} \epsfig{file=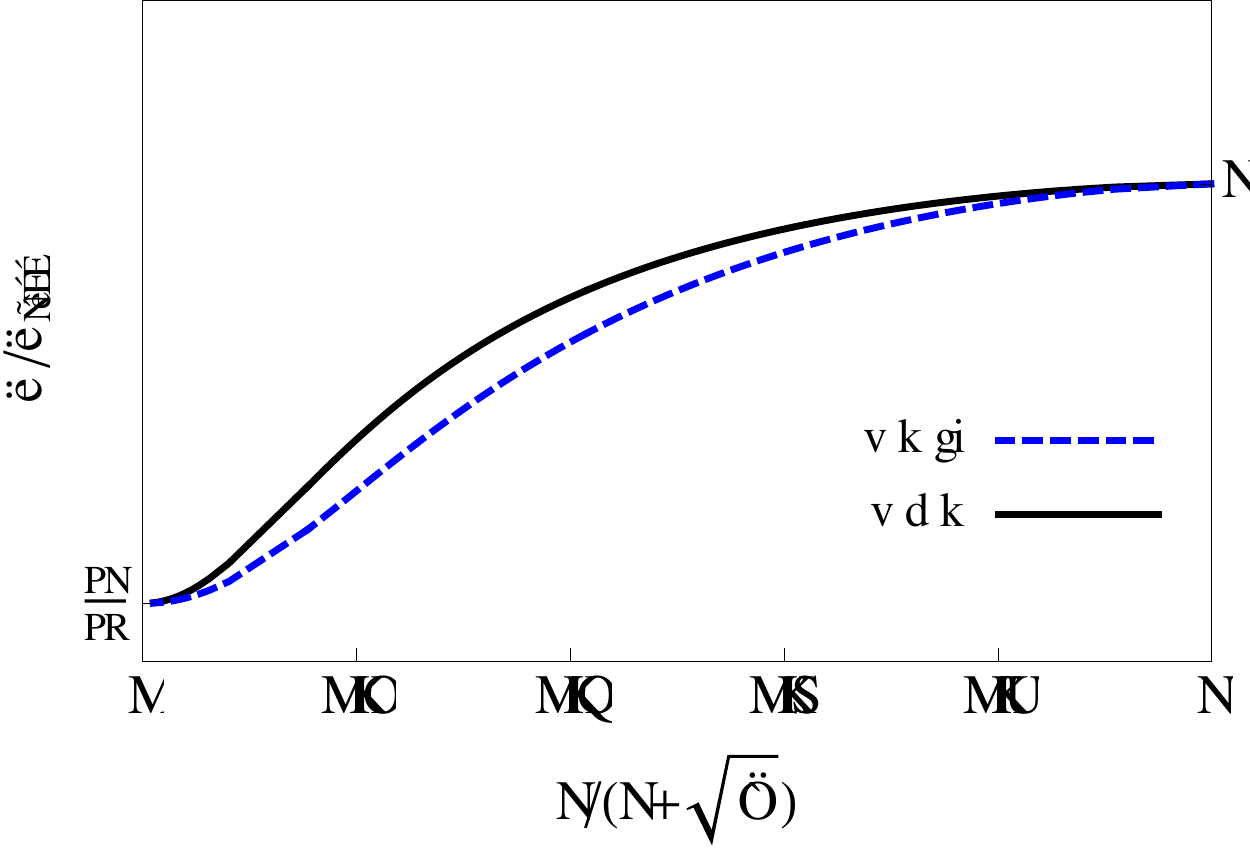,width=0.45\linewidth,angle=0}}
\caption{The  strong-weak entropy density ratio, $s/s_{free}$, as a function of the quantity $1/(1+\sqrt{{\hat g}})$ which  ranges from 0 (strong coupling or low $T$ limit) to 1  (weak coupling or high $T$ limit). The left panel regards the YGN model with $N_b=1$  for $N_f=0,1/4$ and 3 showing that, as ${\hat g} \to \infty$, the curves go to 4/5 ($N_f=0$), 31/35 ($N_f=1/4$) and 49/50 ($N_f=3$). The right panel compare the YGN with at $N_b=1$ and $N_f=1/4$ with the YNJL at $N_b=1$ and $N_f=1/4$. }
\label{Fig2}
\end{figure}

The ratio $s/s_{free}$ can be readily studied at the two extremum limits by using $m_B= 2 \ln \Phi$ and $m_F=0$ for strong $\hat g$ and $m_B=m_F\equiv 0$ for weak $\hat g$. As expected within the weak regime the Stefan-Boltzmann limit is achieved yielding $s/s_{free}=1$. At the strong limit one obtains, after some little algebra, the   result for the YGN/YNJL ratio at infinite coupling
\begin{equation} 
\frac{s}{s_{free}} = \frac {4/5 + 3 N_f/N_b}{1+3N_f/N_b} \;.
\label{ssYGN}
\end{equation}
When $N_f=0$ this relation reproduces the result $s/s_{free} = 4/5$ which has been originally obtained in Ref. \cite{Romatschke:2019ybu} in the case of the scalar $O(N)$ model. As $N_f \to \infty$ the ratio becomes $s/s_{free} \to 1$. Regarding the SUSY theory with cubic superpotential the cases ($N_b=1,N_f = 1/4$)  and  ($N_b=2,N_f = 1/2$), respectively concerning  the YGN and YNJL models,  are the relevant ones.   In this case one obtains $s/s_{free}=31/35$ which is exactly the ratio found within the  $O(N)$ Wess-Zumino theory with a quartic superpotential \cite {DeWolfe:2019etx}. For completeness it is worth recalling that for this model the analytical result quoted in Ref. \cite {DeWolfe:2019etx} is
\begin{equation}
\frac{s}{s_{free}} = \frac{ 4/5 + 3F/(4B)}{1+3F/(4B)} \;,
\label{ssWZ}
\end{equation}
so that when $B=F$ the ratio predicted for the  $O(N)$ ZM model agrees  with the one predicted by the  YGN and YNJL models at ($N_b=1,N_f=1/4$) and ($N_b=1,N_f=1/4$) as Eq. (\ref{ssYGN}) implies. Fig. \ref{Fig2} shows the YGN/YNJL $s/s_{free}$ ratio for all couplings and different values of $N_f$. A detailed discussion about the type of fractionalization implied by Eqs. (\ref{ssYGN}) and (\ref {ssWZ}) can be found in Ref. \cite {Romatschke:2019mjm}.

\begin{figure}[htb]
\centerline{ \epsfig{file=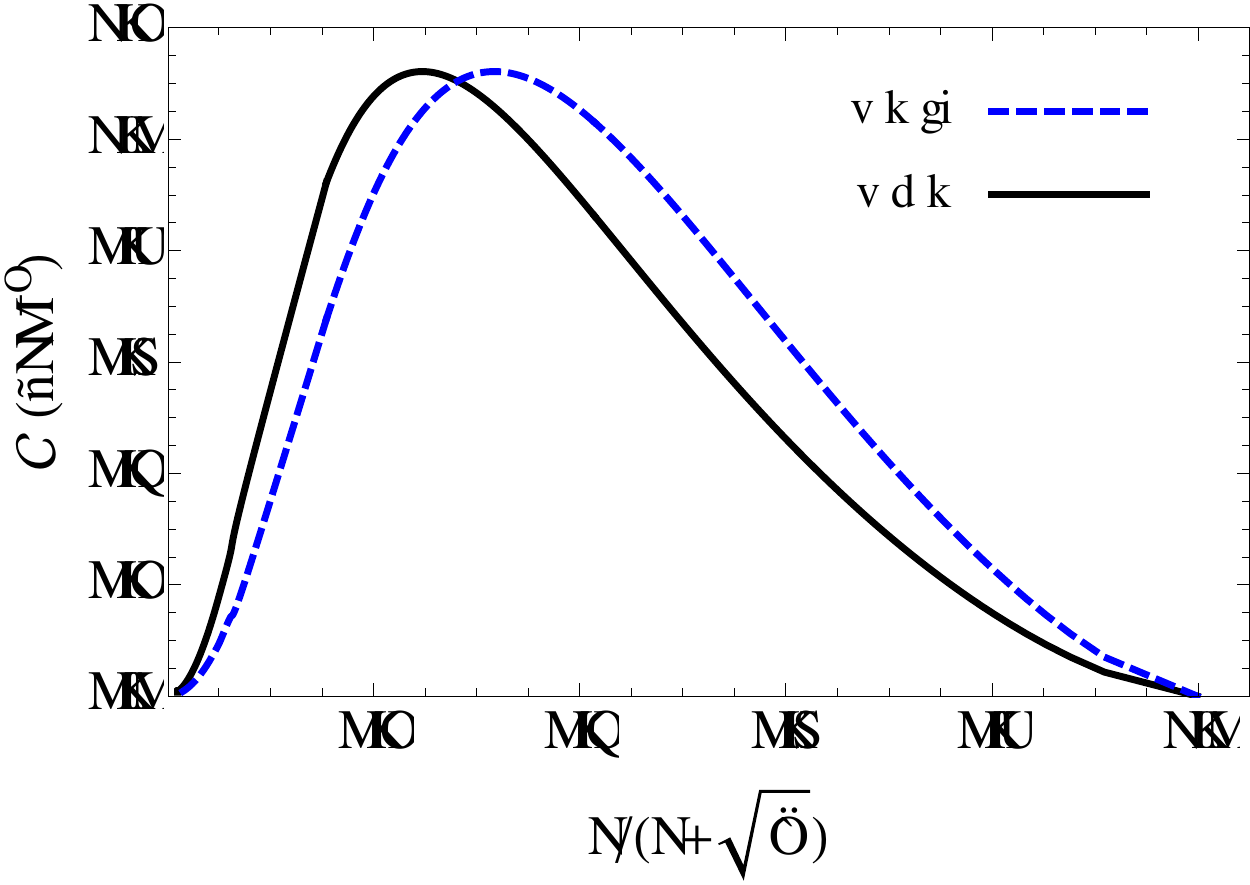,width=0.45\linewidth,angle=0}}
\caption{The conformal measure, $\cal C$, as a function of the quantity $1/(1+\sqrt{{\hat g}})$ which  ranges from 0 (strong coupling or low $T$ limit) to 1  (weak coupling or high $T$ limit). The results are for the YGN with at $N_b=1$ and $N_f=1/4$ and  the YNJL at $N_b=1$ and $N_f=1/4$. }
\label{Fig3}
\end{figure}

\begin{figure}[htb]
\centerline{ \epsfig{file=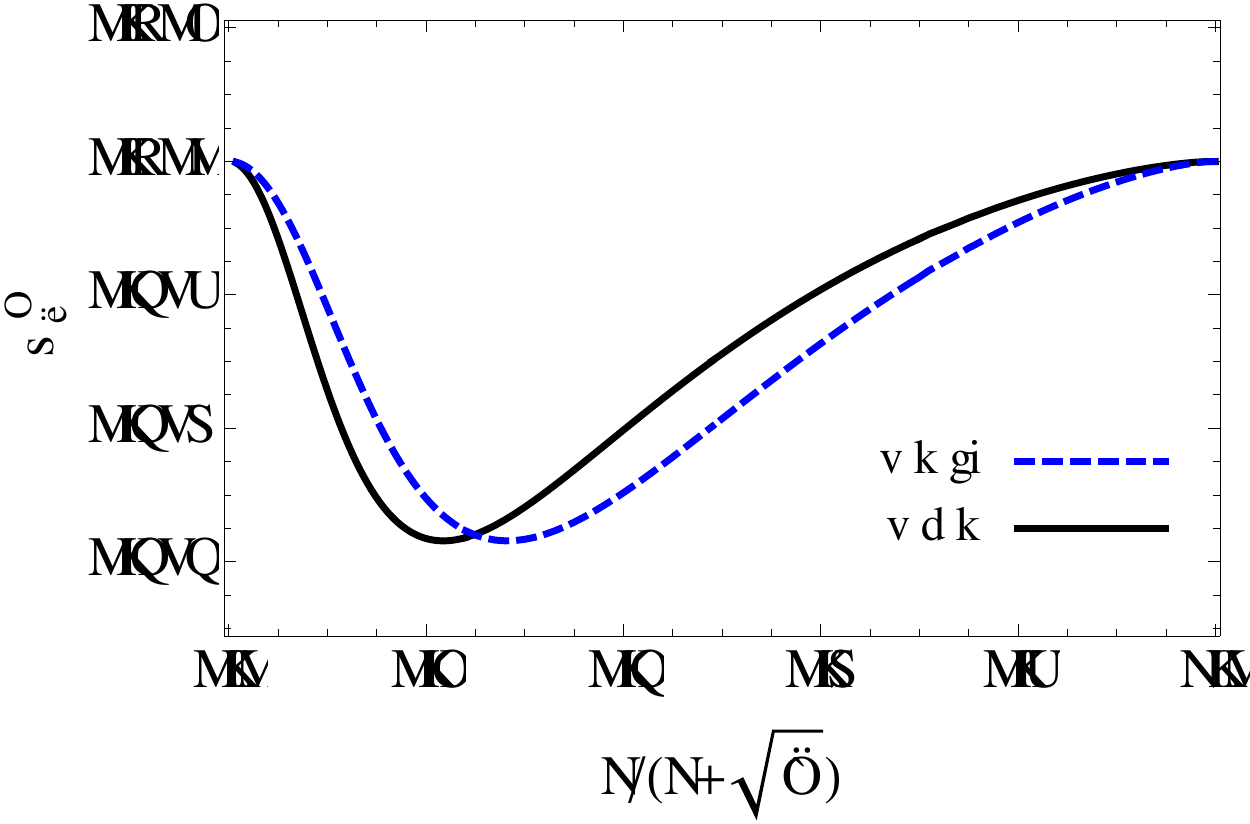,width=0.45\linewidth,angle=0}}
\caption{The speed of sound squared, $V_s^2$, as a function of $(1+\sqrt{{\hat g}})^{-1}$. The results are for the YGN with at $N_b=1$ and $N_f=1/4$ and  the YNJL at $N_b=1$ and $N_f=1/4$. The free gas value is  $V_s^2=0.5$. }
\label{Fig4}
\end{figure}

To examine the conformal measure one can start by analytically obtaining the interaction measure $\Delta =  {\cal E} - 2P = s T - 3 P$. Using $m_F=0$ and $m_B$ as given in Eq. (\ref{mB}) one gets
\begin{equation}
\frac{\Delta}{T^3}=  \frac { m_B^4}{2 {\hat g} } \;,
\label{Delta}
\end{equation}
which, in view of Eq. (\ref{mB}), shows that the YGN/YNJL are CFTs at ${\hat g} =0$ ($m_B=0$) and ${\hat g}=\infty$ ($m_B=2\ln \Phi$) but not in between as illustrated in Fig. \ref {Fig3} which shows ${\cal C}=\Delta/{\cal E}$ for all couplings. The maxima $({\cal C} \simeq 0.011)$ occur at ${\hat g} = 4.63$ for the YNJL  and at twice this value, ${\hat g} = 9.26$, for the YGN which respectively correspond to $(1+\sqrt{{\hat g}})^{-1}\simeq 0.32$ and $(1+\sqrt{{\hat g}})^{-1}\simeq 0.25$. Note also that since Eq. (\ref{Delta})  does not depend on the fermionic degrees of freedom the interaction measure in the YGN/YNJL case is similar to the one found in the  $O(N)$ scalar case with quartic self interactions \cite {Romatschke:2019ybu}.  
From the phenomenological point of view the speed of sound  represents an interesting physical observable to be analyzed at  all coupling values. 
For this purpose Fig. \ref {Fig4}  shows $V_s^2$ as a function of $(1+\sqrt{{\hat g}})^{-1}$ for  the YGN and the YNJL cases indicating that the maximum deviation  from the free gas value, $V_s^2 = 0.5$, occurs at $V_s^2 \simeq 0.494$. It is tempting to interprete this rather small difference as a suggestion that apart from being exact CFTs at ${\hat g}=0$ and ${\hat g}=\infty$ these theories   behave as such, to a good approximation, also at intermediate couplings. 

\begin{figure}[htb]
\centerline{ \epsfig{file=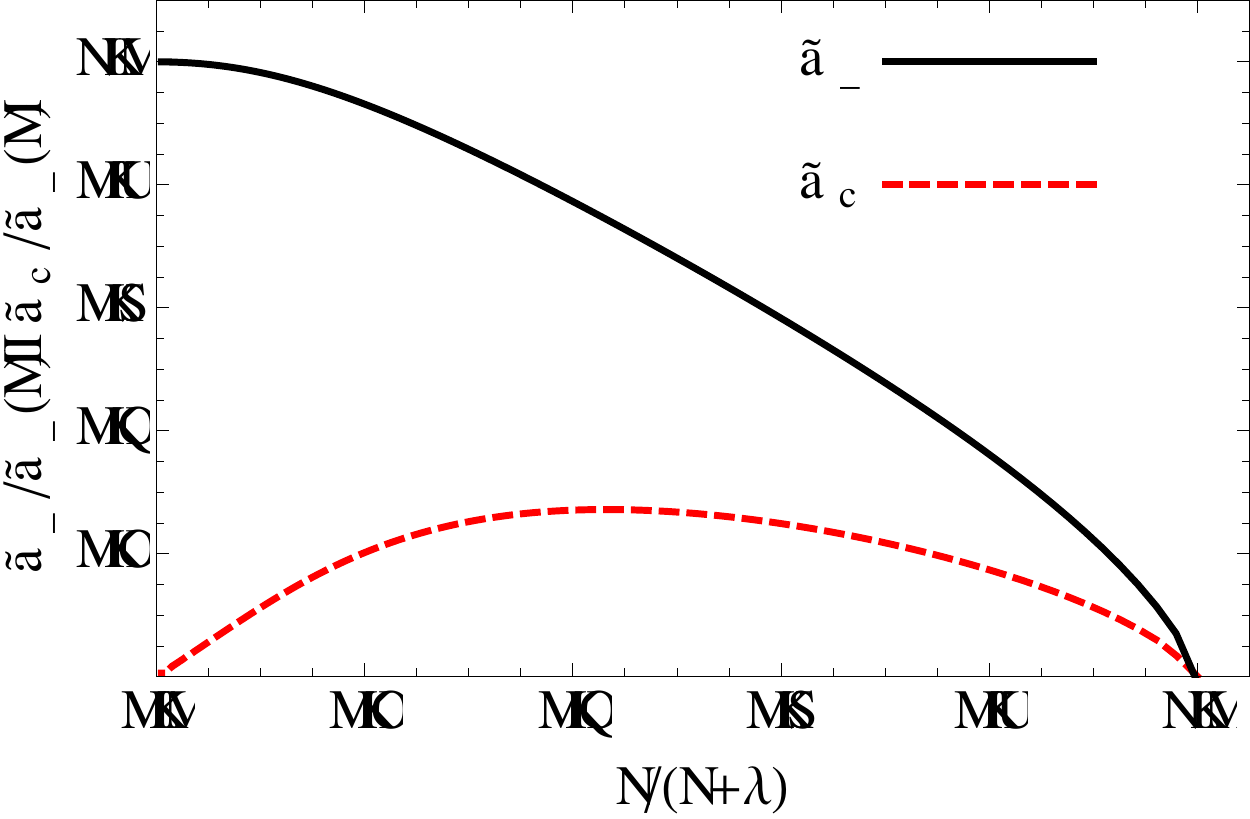,width=0.45\linewidth,angle=0}}
\caption{The bosonic and fermionic masses for the $O(N)$ WZ model, both normalized by $m_B(0)=2 \ln \Phi$, as a function of the quantity $1/(1+\lambda)$ which  ranges from 0 
(strong coupling) to 1  (weak coupling). Taken from Ref. \cite {DeWolfe:2019etx}.}
\label{Fig5}
\end{figure}

\begin{figure}[htb]
\centerline{ \epsfig{file=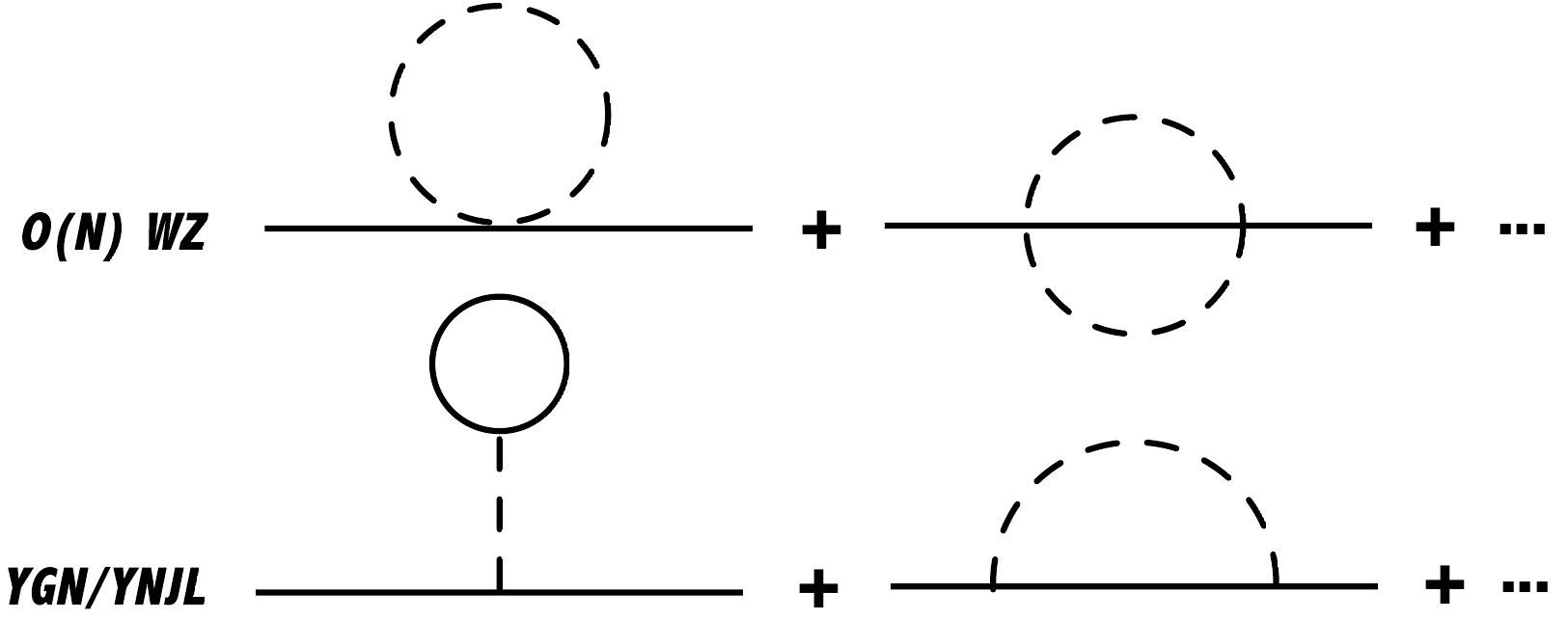,width=0.45\linewidth,angle=0}}
\caption{Feynman diagrams contributing to the fermionic masses, $m_F$, for the different models. Top:  contributions to the O($N$) WZ model with quartic superpotential. The first (scalar tadpole) diagram is the only one contributing in the large-$N$ approximation considered in Ref.\cite {DeWolfe:2019etx} while the second would bring a finite $N$ correction. Bottom: contributions to the YGN/YNJL models (and related WZ with cubic superpotential). The first diagram (fermion tadpole), which vanishes, is the only one consiedered within the MFA adopted here while the second represents an exchange (Fock) type of correction. In both cases the dashed lines represent bosons and continuous lines represent fermions.}
\label{Fig6}
\end{figure}

Finally, in order to better  understand the differences between the $O(N)$ WZ  and the YGN/YNJL (and related WZ with cubic superpotential) models observed at intermediate couplings it is instructive to compared the mass behavior of the former theory with the results displayed in Fig. \ref {Fig1} for the YGN/YNJL case. 
With this aim let us examine Fig. \ref {Fig5} (taken from Ref. \cite {DeWolfe:2019etx}) which displays $m_F$ and $m_B$ as a function of the coupling. The figure shows  that both masses vanish at weak coupling while $m_F \to 0$ and $m_B \to 2 \ln \Phi$ at the strong coupling limit. These two situations coincide with the results for the YGN/YNJL models which also predict ${\cal C}=0$ at these two coupling limits. However, the most important difference arises at intermediate couplings when $m_F$ is non-zero while $m_B$ displays a  behavior which is reminiscent of the one observed in the YGN/YNJL case (compare with Fig. \ref {Fig1}). The origin of the difference can be traced back to the polynomial structure of the potential energy density describing each theory. In the case of the YGN/YNJL the Yukawa vertex is trilinear so that the effective fermionic mass at  the MFA level is given by a (one loop) fermionic tadpole, see right hand side of Eq. (\ref {massaF}) and Fig. \ref {Fig6}, which does not effectively contribute at any coupling. On the other hand in the case of the $O(N)$ the Yukawa vertex if quartic so that the effective fermionic mass at the large-$N$ (one loop) level is given by a (one loop) scalar tadpole which contributes at intermediate couplings, see gap equations in Ref. \cite {DeWolfe:2019etx} and Fig. \ref{Fig6} in the present work.

\section{Conclusions}
The thermodynamics of the massless three dimensional YGN and YNJL models has  been analyzed within the MFA framework for all coupling values. In particular, the results obtained when evaluating the conformal measure, $\cal C$, show that both models behave as CFTs only at infinite and vanishing couplings. Therefore, unlike the $O(N)$ scalar model with sextic interaction or the $O(N)$ WZ model with quartic superpotential, respectively considered in Refs. \cite {Romatschke:2019ybu} and  \cite {DeWolfe:2019etx}, they cannot be considered to represent pure CFTs (for which ${\cal C}$ vanishes at any coupling value). Nevertheless, the results obtained for the entropy density ratio, $s/s_{free}$, at infinite coupling show an exact agreement between the three models when considering  the particular values ($N_f=1/4$, $N_b=1$), ($N_f=1/2$, $N_b=2$), $(F=B)$ for the YGN, YNJL, and $O(N)$ WZ theories respectively. In this strong coupling regime, where all models observe ${\cal C}=0$, one reproduces the ratio $s/s_{free}=31/35$ which was originally obtained in the context of the $O(N)$ WZ model at large-$N$ \cite {DeWolfe:2019etx}. When varying $N_f$ within the YGN (where $N_b=1$) and YNJL (where $N_b=2$) models at infinite coupling one  predicts the $s/s_{free}$ ratio to lie between 4/5 ($N_f=0$) and 1 ($N_f \to \infty$) which is also in agreement with Ref. \cite {DeWolfe:2019etx}. Here, a possible explanation for such an exact agreement was found by examining how the dimensionless fermionic and bosonic masses ($m_F$ and $m_B$) behave at infinite couplings since in all three models $m_F=0$ while $m_B = 2 \ln \Phi$. This allows us to conclude that in this particular regime all these theories display an universal behavior effectively behaving as a gas of massive self interacting bosons plus an independent gas of {\it free} massless fermions. Therefore, when the system is dominated by  fermionic degrees of freedom  ($N_f, F \to \infty$) one obtains $s/s_{free}=1$ whereas in the case of a purely bosonic system ($N_f=F\equiv 0$) the result $s/s_{free}=4/5$, originally obtained in the $O(N)$ scalar model context \cite {Romatschke:2019ybu}, is exactly reproduced.  As  expected, at  vanishing couplings all models behave as a system composed by a gas of massless free bosons plus an independent gas of massless free fermions so that the theories display an universal behavior with ${\cal C}=0$ and $s/s_{free}=1$. The main difference between the YGN/YNJL (which are related to the  WZ with cubic superpotential \cite {Fei:2016sgs}) and the $O(N)$ WZ theories with quartic superpotentials happens at intermediate couplings where the former do not represent CFTs. Based on the present results one may conjecture that one of the main reasons for this difference is the fact that the fermion masses behave in a much  less universal way than  the bosonic masses in the different cases. In particular, within the YGN/YNJL models with trilinear Yukawa vertex the solutions to the gap equations imply that $m_F$ vanishes for all couplings and therefore,  at least within the MFA employed here, the system always behaves as a gas of massive self interacting bosons plus an independent gas of {\it free} massless fermions. In this case the (trilinear) Yukawa coupling ($g_1$) does not play any role and the dynamics is driven solely by the (scalar) quartic coupling ($g_2$) so that the YGN/YNJL and the scalar $O(N)$ model with quartic interaction studied in Ref. \cite{Romatschke:2019ybu} display a similar conformal measure. On the other hand, within the $O(N)$ WZ theory with quartic Yukawa vertex,  $m_F$ attains finite values at intermediate couplings while vanishing only at the extremum $\lambda=0$ and $\lambda= \infty$ values. In summary, the results obtained here together with the ones obtained in Refs. \cite {Romatschke:2019ybu} and \cite {DeWolfe:2019etx}, confirm that the $O(N)$ scalar model with sextic interaction and $O(N)$ WZ model with quartic superpotential represent pure CFTs in contrast to the $O(N)$ scalar model with quartic interaction and the YGN/YNJL models (as well as the related WZ model with cubic superpotential) which behave as CFTs only at vanishing and infinite couplings where the thermodynamical behavior displayed by all theories appears to be more universal. However, from a more quantitative point of view it is worth recalling that the  values reached by the of speed of sound within the  YGN/YNJL models at intermediate couplings are never greater than $V_s^2\simeq 0.494$ which is still very close to the free gas value, $V_s^2\simeq 0.5$, observed by pure CFTs. 
Regarding further refinements one question that immediately arises regards the reliability of all those results which were obtained with the MFA, in the present work, and at large-$N$  in Refs. \cite {Romatschke:2019ybu, DeWolfe:2019etx}. This becomes a very relevant question especially if one recalls how these two approximations may fail in correctly describing the thermodynamics of low dimensional systems at finite temperatures. One example occurs within the related Gross-Neveu model in $2+1d$ where the large-$N$ approximation  predicts that chiral symmetry at finite temperatures and densities is restored through a second order phase transition at all finite temperatures and through a first order transition only at $T=0$ \cite  {Klimenko:1987gi,*Rosenstein:1988dj,*Rosenstein:1988pt}. In this situation the inclusion of finite $N$ effects \cite {Kneur:2007vm, *Kneur:2007vj} changes the transition pattern predicting that a first order transition boundary, also present at low finite temperatures, terminates at a tricritical point located at intermediate temperatures and densities (missed by the large-$N$ approximation) in accordance with  lattice  predictions \cite {Kogut:1999um} (see Ref.\cite {Kneur:2007vm, *Kneur:2007vj} for more examples).  One possiblity   to improve the MFA and large-$N$ evaluations is to consider alternative non perturbative techniques such as the optimized perturbation theory \cite {Okopinska:1987hp,*Duncan:1988hw},  used in Ref.\cite {Kneur:2007vm, *Kneur:2007vj}, or the resummation scheme recently proposed in Ref. \cite {Romatschke:2019rjk} so as to dress the fermionic masses with  exchange (Fock like) type of contributions which are not considered at the MFA/large-$N$/Hartree level.

\acknowledgments
I would like to thank  Paul Romatschke for sharing his knowledge on this topic  as well as for offering suggestions regarding the manuscript. I also thank Simone Giombi, Jean-Lo\"{\i}c Kneur, Odilon Louren\c co, and Rudnei Ramos for discussions related to this work. The author is  partially supported by Conselho Nacional de Desenvolvimento Cient\'{\i}fico e Tecnol\'{o}gico (CNPq-Brazil), Process No 303846/2017-8. This work has also been financed  in  part  by    INCT-FNA, Process No.  464898/2014-5. 
 
\bibliographystyle{apsrev4-1}
\bibliography{ref_marcus} 

\begin{thebibliography}{32}%
\makeatletter
\providecommand \@ifxundefined [1]{%
 \@ifx{#1\undefined}
}%
\providecommand \@ifnum [1]{%
 \ifnum #1\expandafter \@firstoftwo
 \else \expandafter \@secondoftwo
 \fi
}%
\providecommand \@ifx [1]{%
 \ifx #1\expandafter \@firstoftwo
 \else \expandafter \@secondoftwo
 \fi
}%
\providecommand \natexlab [1]{#1}%
\providecommand \enquote  [1]{``#1''}%
\providecommand \bibnamefont  [1]{#1}%
\providecommand \bibfnamefont [1]{#1}%
\providecommand \citenamefont [1]{#1}%
\providecommand \href@noop [0]{\@secondoftwo}%
\providecommand \href [0]{\begingroup \@sanitize@url \@href}%
\providecommand \@href[1]{\@@startlink{#1}\@@href}%
\providecommand \@@href[1]{\endgroup#1\@@endlink}%
\providecommand \@sanitize@url [0]{\catcode `\\12\catcode `\$12\catcode
  `\&12\catcode `\#12\catcode `\^12\catcode `\_12\catcode `\%12\relax}%
\providecommand \@@startlink[1]{}%
\providecommand \@@endlink[0]{}%
\providecommand \url  [0]{\begingroup\@sanitize@url \@url }%
\providecommand \@url [1]{\endgroup\@href {#1}{\urlprefix }}%
\providecommand \urlprefix  [0]{URL }%
\providecommand \Eprint [0]{\href }%
\providecommand \doibase [0]{http://dx.doi.org/}%
\providecommand \selectlanguage [0]{\@gobble}%
\providecommand \bibinfo  [0]{\@secondoftwo}%
\providecommand \bibfield  [0]{\@secondoftwo}%
\providecommand \translation [1]{[#1]}%
\providecommand \BibitemOpen [0]{}%
\providecommand \bibitemStop [0]{}%
\providecommand \bibitemNoStop [0]{.\EOS\space}%
\providecommand \EOS [0]{\spacefactor3000\relax}%
\providecommand \BibitemShut  [1]{\csname bibitem#1\endcsname}%
\let\auto@bib@innerbib\@empty
\bibitem [{\citenamefont
  {Romatschke}(2019{\natexlab{a}})}]{Romatschke:2019ybu}%
  \BibitemOpen
  \bibfield  {author} {\bibinfo {author} {\bibfnamefont {P.}~\bibnamefont
  {Romatschke}},\ }\href {\doibase 10.1103/PhysRevLett.122.231603} {\bibfield
  {journal} {\bibinfo  {journal} {Phys. Rev. Lett.}\ }\textbf {\bibinfo
  {volume} {122}},\ \bibinfo {pages} {231603} (\bibinfo {year}
  {2019}{\natexlab{a}})},\ \bibinfo {note} {[Erratum: Phys.Rev.Lett. 123,
  209901 (2019)]},\ \Eprint {http://arxiv.org/abs/1904.09995} {arXiv:1904.09995
  [hep-th]} \BibitemShut {NoStop}%
\bibitem [{\citenamefont {Gubser}\ \emph {et~al.}(1998)\citenamefont {Gubser},
  \citenamefont {Klebanov},\ and\ \citenamefont {Tseytlin}}]{Gubser:1998nz}%
  \BibitemOpen
  \bibfield  {author} {\bibinfo {author} {\bibfnamefont {S.~S.}\ \bibnamefont
  {Gubser}}, \bibinfo {author} {\bibfnamefont {I.~R.}\ \bibnamefont
  {Klebanov}}, \ and\ \bibinfo {author} {\bibfnamefont {A.~A.}\ \bibnamefont
  {Tseytlin}},\ }\href {\doibase 10.1016/S0550-3213(98)00514-8} {\bibfield
  {journal} {\bibinfo  {journal} {Nucl. Phys. B}\ }\textbf {\bibinfo {volume}
  {534}},\ \bibinfo {pages} {202} (\bibinfo {year} {1998})},\ \Eprint
  {http://arxiv.org/abs/hep-th/9805156} {arXiv:hep-th/9805156} \BibitemShut
  {NoStop}%
\bibitem [{\citenamefont {DeWolfe}\ and\ \citenamefont
  {Romatschke}(2019)}]{DeWolfe:2019etx}%
  \BibitemOpen
  \bibfield  {author} {\bibinfo {author} {\bibfnamefont {O.}~\bibnamefont
  {DeWolfe}}\ and\ \bibinfo {author} {\bibfnamefont {P.}~\bibnamefont
  {Romatschke}},\ }\href {\doibase 10.1007/JHEP10(2019)272} {\bibfield
  {journal} {\bibinfo  {journal} {JHEP}\ }\textbf {\bibinfo {volume} {10}},\
  \bibinfo {pages} {272} (\bibinfo {year} {2019})},\ \Eprint
  {http://arxiv.org/abs/1905.06355} {arXiv:1905.06355 [hep-th]} \BibitemShut
  {NoStop}%
\bibitem [{\citenamefont {Wess}\ and\ \citenamefont
  {Zumino}(1974)}]{Wess:1974tw}%
  \BibitemOpen
  \bibfield  {author} {\bibinfo {author} {\bibfnamefont {J.}~\bibnamefont
  {Wess}}\ and\ \bibinfo {author} {\bibfnamefont {B.}~\bibnamefont {Zumino}},\
  }\href {\doibase 10.1016/0550-3213(74)90355-1} {\bibfield  {journal}
  {\bibinfo  {journal} {Nucl. Phys. B}\ }\textbf {\bibinfo {volume} {70}},\
  \bibinfo {pages} {39} (\bibinfo {year} {1974})}\BibitemShut {NoStop}%
\bibitem [{\citenamefont {Hasenfratz}\ \emph {et~al.}(1991)\citenamefont
  {Hasenfratz}, \citenamefont {Hasenfratz}, \citenamefont {Jansen},
  \citenamefont {Kuti},\ and\ \citenamefont {Shen}}]{Hasenfratz:1991it}%
  \BibitemOpen
  \bibfield  {author} {\bibinfo {author} {\bibfnamefont {A.}~\bibnamefont
  {Hasenfratz}}, \bibinfo {author} {\bibfnamefont {P.}~\bibnamefont
  {Hasenfratz}}, \bibinfo {author} {\bibfnamefont {K.}~\bibnamefont {Jansen}},
  \bibinfo {author} {\bibfnamefont {J.}~\bibnamefont {Kuti}}, \ and\ \bibinfo
  {author} {\bibfnamefont {Y.}~\bibnamefont {Shen}},\ }\href {\doibase
  10.1016/0550-3213(91)90607-Y} {\bibfield  {journal} {\bibinfo  {journal}
  {Nucl. Phys. B}\ }\textbf {\bibinfo {volume} {365}},\ \bibinfo {pages} {79}
  (\bibinfo {year} {1991})}\BibitemShut {NoStop}%
\bibitem [{\citenamefont {Zinn-Justin}(1991)}]{ZinnJustin:1991yn}%
  \BibitemOpen
  \bibfield  {author} {\bibinfo {author} {\bibfnamefont {J.}~\bibnamefont
  {Zinn-Justin}},\ }\href {\doibase 10.1016/0550-3213(91)90043-W} {\bibfield
  {journal} {\bibinfo  {journal} {Nucl. Phys. B}\ }\textbf {\bibinfo {volume}
  {367}},\ \bibinfo {pages} {105} (\bibinfo {year} {1991})}\BibitemShut
  {NoStop}%
\bibitem [{\citenamefont {Walecka}(1974)}]{Walecka:1974qa}%
  \BibitemOpen
  \bibfield  {author} {\bibinfo {author} {\bibfnamefont {J.}~\bibnamefont
  {Walecka}},\ }\href {\doibase 10.1016/0003-4916(74)90208-5} {\bibfield
  {journal} {\bibinfo  {journal} {Annals Phys.}\ }\textbf {\bibinfo {volume}
  {83}},\ \bibinfo {pages} {491} (\bibinfo {year} {1974})}\BibitemShut
  {NoStop}%
\bibitem [{\citenamefont {Serot}\ and\ \citenamefont
  {Walecka}(1986)}]{Serot:1984ey}%
  \BibitemOpen
  \bibfield  {author} {\bibinfo {author} {\bibfnamefont {B.~D.}\ \bibnamefont
  {Serot}}\ and\ \bibinfo {author} {\bibfnamefont {J.~D.}\ \bibnamefont
  {Walecka}},\ }\href@noop {} {\bibfield  {journal} {\bibinfo  {journal} {Adv.
  Nucl. Phys.}\ }\textbf {\bibinfo {volume} {16}},\ \bibinfo {pages} {1}
  (\bibinfo {year} {1986})}\BibitemShut {NoStop}%
\bibitem [{\citenamefont {Gell-Mann}\ and\ \citenamefont
  {Levy}(1960)}]{GellMann:1960np}%
  \BibitemOpen
  \bibfield  {author} {\bibinfo {author} {\bibfnamefont {M.}~\bibnamefont
  {Gell-Mann}}\ and\ \bibinfo {author} {\bibfnamefont {M.}~\bibnamefont
  {Levy}},\ }\href {\doibase 10.1007/BF02859738} {\bibfield  {journal}
  {\bibinfo  {journal} {Nuovo Cim.}\ }\textbf {\bibinfo {volume} {16}},\
  \bibinfo {pages} {705} (\bibinfo {year} {1960})}\BibitemShut {NoStop}%
\bibitem [{\citenamefont {Gross}\ and\ \citenamefont
  {Neveu}(1974)}]{Gross:1974jv}%
  \BibitemOpen
  \bibfield  {author} {\bibinfo {author} {\bibfnamefont {D.~J.}\ \bibnamefont
  {Gross}}\ and\ \bibinfo {author} {\bibfnamefont {A.}~\bibnamefont {Neveu}},\
  }\href {\doibase 10.1103/PhysRevD.10.3235} {\bibfield  {journal} {\bibinfo
  {journal} {Phys. Rev. D}\ }\textbf {\bibinfo {volume} {10}},\ \bibinfo
  {pages} {3235} (\bibinfo {year} {1974})}\BibitemShut {NoStop}%
\bibitem [{\citenamefont {Nambu}\ and\ \citenamefont
  {Jona-Lasinio}(1961{\natexlab{a}})}]{Nambu:1961tp}%
  \BibitemOpen
  \bibfield  {author} {\bibinfo {author} {\bibfnamefont {Y.}~\bibnamefont
  {Nambu}}\ and\ \bibinfo {author} {\bibfnamefont {G.}~\bibnamefont
  {Jona-Lasinio}},\ }\href {\doibase 10.1103/PhysRev.122.345} {\bibfield
  {journal} {\bibinfo  {journal} {Phys. Rev.}\ }\textbf {\bibinfo {volume}
  {122}},\ \bibinfo {pages} {345} (\bibinfo {year}
  {1961}{\natexlab{a}})}\BibitemShut {NoStop}%
\bibitem [{\citenamefont {Nambu}\ and\ \citenamefont
  {Jona-Lasinio}(1961{\natexlab{b}})}]{Nambu:1961fr}%
  \BibitemOpen
  \bibfield  {author} {\bibinfo {author} {\bibfnamefont {Y.}~\bibnamefont
  {Nambu}}\ and\ \bibinfo {author} {\bibfnamefont {G.}~\bibnamefont
  {Jona-Lasinio}},\ }\href {\doibase 10.1103/PhysRev.124.246} {\bibfield
  {journal} {\bibinfo  {journal} {Phys. Rev.}\ }\textbf {\bibinfo {volume}
  {124}},\ \bibinfo {pages} {246} (\bibinfo {year}
  {1961}{\natexlab{b}})}\BibitemShut {NoStop}%
\bibitem [{\citenamefont {Fei}\ \emph {et~al.}(2016)\citenamefont {Fei},
  \citenamefont {Giombi}, \citenamefont {Klebanov},\ and\ \citenamefont
  {Tarnopolsky}}]{Fei:2016sgs}%
  \BibitemOpen
  \bibfield  {author} {\bibinfo {author} {\bibfnamefont {L.}~\bibnamefont
  {Fei}}, \bibinfo {author} {\bibfnamefont {S.}~\bibnamefont {Giombi}},
  \bibinfo {author} {\bibfnamefont {I.~R.}\ \bibnamefont {Klebanov}}, \ and\
  \bibinfo {author} {\bibfnamefont {G.}~\bibnamefont {Tarnopolsky}},\ }\href
  {\doibase 10.1093/ptep/ptw120} {\bibfield  {journal} {\bibinfo  {journal}
  {PTEP}\ }\textbf {\bibinfo {volume} {2016}},\ \bibinfo {pages} {12C105}
  (\bibinfo {year} {2016})},\ \Eprint {http://arxiv.org/abs/1607.05316}
  {arXiv:1607.05316 [hep-th]} \BibitemShut {NoStop}%
\bibitem [{\citenamefont {Appelquist}\ \emph {et~al.}(1986)\citenamefont
  {Appelquist}, \citenamefont {Bowick}, \citenamefont {Karabali},\ and\
  \citenamefont {Wijewardhana}}]{Appelquist:1986fd}%
  \BibitemOpen
  \bibfield  {author} {\bibinfo {author} {\bibfnamefont {T.~W.}\ \bibnamefont
  {Appelquist}}, \bibinfo {author} {\bibfnamefont {M.~J.}\ \bibnamefont
  {Bowick}}, \bibinfo {author} {\bibfnamefont {D.}~\bibnamefont {Karabali}}, \
  and\ \bibinfo {author} {\bibfnamefont {L.}~\bibnamefont {Wijewardhana}},\
  }\href {\doibase 10.1103/PhysRevD.33.3704} {\bibfield  {journal} {\bibinfo
  {journal} {Phys. Rev. D}\ }\textbf {\bibinfo {volume} {33}},\ \bibinfo
  {pages} {3704} (\bibinfo {year} {1986})}\BibitemShut {NoStop}%
\bibitem [{\citenamefont {Rosenstein}\ \emph {et~al.}(1991)\citenamefont
  {Rosenstein}, \citenamefont {Warr},\ and\ \citenamefont
  {Park}}]{Rosenstein:1990nm}%
  \BibitemOpen
  \bibfield  {author} {\bibinfo {author} {\bibfnamefont {B.}~\bibnamefont
  {Rosenstein}}, \bibinfo {author} {\bibfnamefont {B.}~\bibnamefont {Warr}}, \
  and\ \bibinfo {author} {\bibfnamefont {S.}~\bibnamefont {Park}},\ }\href
  {\doibase 10.1016/0370-1573(91)90129-A} {\bibfield  {journal} {\bibinfo
  {journal} {Phys. Rept.}\ }\textbf {\bibinfo {volume} {205}},\ \bibinfo
  {pages} {59} (\bibinfo {year} {1991})}\BibitemShut {NoStop}%
\bibitem [{\citenamefont {Bailin}\ and\ \citenamefont {Love}(1993)}]{Bailin}%
  \BibitemOpen
  \bibfield  {author} {\bibinfo {author} {\bibfnamefont {D.}~\bibnamefont
  {Bailin}}\ and\ \bibinfo {author} {\bibfnamefont {A.}~\bibnamefont {Love}},\
  }\href@noop {} {\emph {\bibinfo {title} {Introduction to Gauge Field
  Theory}}},\ edited by\ \bibinfo {editor} {\bibfnamefont {D.~F.}\ \bibnamefont
  {Brewer}}\ (\bibinfo  {publisher} {Taylor \& Francis Group},\ \bibinfo {year}
  {1993})\BibitemShut {NoStop}%
\bibitem [{\citenamefont {Laine}\ and\ \citenamefont
  {Vuorinen}(2016)}]{Laine:2016hma}%
  \BibitemOpen
  \bibfield  {author} {\bibinfo {author} {\bibfnamefont {M.}~\bibnamefont
  {Laine}}\ and\ \bibinfo {author} {\bibfnamefont {A.}~\bibnamefont
  {Vuorinen}},\ }\href {\doibase 10.1007/978-3-319-31933-9} {\emph {\bibinfo
  {title} {{Basics of Thermal Field Theory}}}},\ Vol.\ \bibinfo {volume} {925}\
  (\bibinfo  {publisher} {Springer},\ \bibinfo {year} {2016})\ \Eprint
  {http://arxiv.org/abs/1701.01554} {arXiv:1701.01554 [hep-ph]} \BibitemShut
  {NoStop}%
\bibitem [{\citenamefont {Kapusta}\ and\ \citenamefont
  {Gale}(2011)}]{Kapusta:2006pm}%
  \BibitemOpen
  \bibfield  {author} {\bibinfo {author} {\bibfnamefont {J.}~\bibnamefont
  {Kapusta}}\ and\ \bibinfo {author} {\bibfnamefont {C.}~\bibnamefont {Gale}},\
  }\href {\doibase 10.1017/CBO9780511535130} {\emph {\bibinfo {title}
  {{Finite-temperature field theory: Principles and applications}}}},\
  Cambridge Monographs on Mathematical Physics\ (\bibinfo  {publisher}
  {Cambridge University Press},\ \bibinfo {year} {2011})\BibitemShut {NoStop}%
\bibitem [{\citenamefont {Iliesiu}\ \emph {et~al.}(2016)\citenamefont
  {Iliesiu}, \citenamefont {Kos}, \citenamefont {Poland}, \citenamefont {Pufu},
  \citenamefont {Simmons-Duffin},\ and\ \citenamefont
  {Yacoby}}]{Iliesiu:2015qra}%
  \BibitemOpen
  \bibfield  {author} {\bibinfo {author} {\bibfnamefont {L.}~\bibnamefont
  {Iliesiu}}, \bibinfo {author} {\bibfnamefont {F.}~\bibnamefont {Kos}},
  \bibinfo {author} {\bibfnamefont {D.}~\bibnamefont {Poland}}, \bibinfo
  {author} {\bibfnamefont {S.~S.}\ \bibnamefont {Pufu}}, \bibinfo {author}
  {\bibfnamefont {D.}~\bibnamefont {Simmons-Duffin}}, \ and\ \bibinfo {author}
  {\bibfnamefont {R.}~\bibnamefont {Yacoby}},\ }\href {\doibase
  10.1007/JHEP03(2016)120} {\bibfield  {journal} {\bibinfo  {journal} {JHEP}\
  }\textbf {\bibinfo {volume} {03}},\ \bibinfo {pages} {120} (\bibinfo {year}
  {2016})},\ \Eprint {http://arxiv.org/abs/1508.00012} {arXiv:1508.00012
  [hep-th]} \BibitemShut {NoStop}%
\bibitem [{\citenamefont {Grover}\ \emph {et~al.}(2014)\citenamefont {Grover},
  \citenamefont {Sheng},\ and\ \citenamefont {Vishwanath}}]{Grover:2013rc}%
  \BibitemOpen
  \bibfield  {author} {\bibinfo {author} {\bibfnamefont {T.}~\bibnamefont
  {Grover}}, \bibinfo {author} {\bibfnamefont {D.}~\bibnamefont {Sheng}}, \
  and\ \bibinfo {author} {\bibfnamefont {A.}~\bibnamefont {Vishwanath}},\
  }\href {\doibase 10.1126/science.1248253} {\bibfield  {journal} {\bibinfo
  {journal} {Science}\ }\textbf {\bibinfo {volume} {344}},\ \bibinfo {pages}
  {280} (\bibinfo {year} {2014})},\ \Eprint {http://arxiv.org/abs/1301.7449}
  {arXiv:1301.7449 [cond-mat.str-el]} \BibitemShut {NoStop}%
\bibitem [{\citenamefont {Bashkirov}(2013)}]{Bashkirov:2013vya}%
  \BibitemOpen
  \bibfield  {author} {\bibinfo {author} {\bibfnamefont {D.}~\bibnamefont
  {Bashkirov}},\ }\href@noop {} {\  (\bibinfo {year} {2013})},\ \Eprint
  {http://arxiv.org/abs/1310.8255} {arXiv:1310.8255 [hep-th]} \BibitemShut
  {NoStop}%
\bibitem [{\citenamefont {Sonoda}(2011)}]{Sonoda:2011qd}%
  \BibitemOpen
  \bibfield  {author} {\bibinfo {author} {\bibfnamefont {H.}~\bibnamefont
  {Sonoda}},\ }\href {\doibase 10.1143/PTP.126.57} {\bibfield  {journal}
  {\bibinfo  {journal} {Prog. Theor. Phys.}\ }\textbf {\bibinfo {volume}
  {126}},\ \bibinfo {pages} {57} (\bibinfo {year} {2011})},\ \Eprint
  {http://arxiv.org/abs/1102.3974} {arXiv:1102.3974 [hep-th]} \BibitemShut
  {NoStop}%
\bibitem [{\citenamefont
  {Romatschke}(2019{\natexlab{b}})}]{Romatschke:2019mjm}%
  \BibitemOpen
  \bibfield  {author} {\bibinfo {author} {\bibfnamefont {P.}~\bibnamefont
  {Romatschke}},\ }\href {\doibase 10.1103/PhysRevLett.123.241602} {\bibfield
  {journal} {\bibinfo  {journal} {Phys. Rev. Lett.}\ }\textbf {\bibinfo
  {volume} {123}},\ \bibinfo {pages} {241602} (\bibinfo {year}
  {2019}{\natexlab{b}})},\ \Eprint {http://arxiv.org/abs/1908.02758}
  {arXiv:1908.02758 [hep-th]} \BibitemShut {NoStop}%
\bibitem [{\citenamefont {Klimenko}(1988)}]{Klimenko:1987gi}%
  \BibitemOpen
  \bibfield  {author} {\bibinfo {author} {\bibfnamefont {K.}~\bibnamefont
  {Klimenko}},\ }\href {\doibase 10.1007/BF01578141} {\bibfield  {journal}
  {\bibinfo  {journal} {Z. Phys. C}\ }\textbf {\bibinfo {volume} {37}},\
  \bibinfo {pages} {457} (\bibinfo {year} {1988})}\BibitemShut {NoStop}%
\bibitem [{\citenamefont {Rosenstein}\ \emph
  {et~al.}(1989{\natexlab{a}})\citenamefont {Rosenstein}, \citenamefont
  {Warr},\ and\ \citenamefont {Park}}]{Rosenstein:1988dj}%
  \BibitemOpen
  \bibfield  {author} {\bibinfo {author} {\bibfnamefont {B.}~\bibnamefont
  {Rosenstein}}, \bibinfo {author} {\bibfnamefont {B.}~\bibnamefont {Warr}}, \
  and\ \bibinfo {author} {\bibfnamefont {S.}~\bibnamefont {Park}},\ }\href
  {\doibase 10.1103/PhysRevD.39.3088} {\bibfield  {journal} {\bibinfo
  {journal} {Phys. Rev. D}\ }\textbf {\bibinfo {volume} {39}},\ \bibinfo
  {pages} {3088} (\bibinfo {year} {1989}{\natexlab{a}})}\BibitemShut {NoStop}%
\bibitem [{\citenamefont {Rosenstein}\ \emph
  {et~al.}(1989{\natexlab{b}})\citenamefont {Rosenstein}, \citenamefont
  {Warr},\ and\ \citenamefont {Park}}]{Rosenstein:1988pt}%
  \BibitemOpen
  \bibfield  {author} {\bibinfo {author} {\bibfnamefont {B.}~\bibnamefont
  {Rosenstein}}, \bibinfo {author} {\bibfnamefont {B.~J.}\ \bibnamefont
  {Warr}}, \ and\ \bibinfo {author} {\bibfnamefont {S.~H.}\ \bibnamefont
  {Park}},\ }\href {\doibase 10.1103/PhysRevLett.62.1433} {\bibfield  {journal}
  {\bibinfo  {journal} {Phys. Rev. Lett.}\ }\textbf {\bibinfo {volume} {62}},\
  \bibinfo {pages} {1433} (\bibinfo {year} {1989}{\natexlab{b}})}\BibitemShut
  {NoStop}%
\bibitem [{\citenamefont {Kneur}\ \emph
  {et~al.}(2007{\natexlab{a}})\citenamefont {Kneur}, \citenamefont {Pinto},
  \citenamefont {Ramos},\ and\ \citenamefont {Staudt}}]{Kneur:2007vm}%
  \BibitemOpen
  \bibfield  {author} {\bibinfo {author} {\bibfnamefont {J.-L.}\ \bibnamefont
  {Kneur}}, \bibinfo {author} {\bibfnamefont {M.~B.}\ \bibnamefont {Pinto}},
  \bibinfo {author} {\bibfnamefont {R.~O.}\ \bibnamefont {Ramos}}, \ and\
  \bibinfo {author} {\bibfnamefont {E.}~\bibnamefont {Staudt}},\ }\href
  {\doibase 10.1103/PhysRevD.76.045020} {\bibfield  {journal} {\bibinfo
  {journal} {Phys. Rev. D}\ }\textbf {\bibinfo {volume} {76}},\ \bibinfo
  {pages} {045020} (\bibinfo {year} {2007}{\natexlab{a}})},\ \Eprint
  {http://arxiv.org/abs/0705.0676} {arXiv:0705.0676 [hep-th]} \BibitemShut
  {NoStop}%
\bibitem [{\citenamefont {Kneur}\ \emph
  {et~al.}(2007{\natexlab{b}})\citenamefont {Kneur}, \citenamefont {Pinto},
  \citenamefont {Ramos},\ and\ \citenamefont {Staudt}}]{Kneur:2007vj}%
  \BibitemOpen
  \bibfield  {author} {\bibinfo {author} {\bibfnamefont {J.-L.}\ \bibnamefont
  {Kneur}}, \bibinfo {author} {\bibfnamefont {M.~B.}\ \bibnamefont {Pinto}},
  \bibinfo {author} {\bibfnamefont {R.~O.}\ \bibnamefont {Ramos}}, \ and\
  \bibinfo {author} {\bibfnamefont {E.}~\bibnamefont {Staudt}},\ }\href
  {\doibase 10.1016/j.physletb.2007.10.013} {\bibfield  {journal} {\bibinfo
  {journal} {Phys. Lett. B}\ }\textbf {\bibinfo {volume} {657}},\ \bibinfo
  {pages} {136} (\bibinfo {year} {2007}{\natexlab{b}})},\ \Eprint
  {http://arxiv.org/abs/0705.0673} {arXiv:0705.0673 [hep-ph]} \BibitemShut
  {NoStop}%
\bibitem [{\citenamefont {Kogut}\ and\ \citenamefont
  {Strouthos}(2001)}]{Kogut:1999um}%
  \BibitemOpen
  \bibfield  {author} {\bibinfo {author} {\bibfnamefont {J.}~\bibnamefont
  {Kogut}}\ and\ \bibinfo {author} {\bibfnamefont {C.}~\bibnamefont
  {Strouthos}},\ }\href {\doibase 10.1103/PhysRevD.63.054502} {\bibfield
  {journal} {\bibinfo  {journal} {Phys. Rev. D}\ }\textbf {\bibinfo {volume}
  {63}},\ \bibinfo {pages} {054502} (\bibinfo {year} {2001})},\ \Eprint
  {http://arxiv.org/abs/hep-lat/9904008} {arXiv:hep-lat/9904008} \BibitemShut
  {NoStop}%
\bibitem [{\citenamefont {Okopinska}(1987)}]{Okopinska:1987hp}%
  \BibitemOpen
  \bibfield  {author} {\bibinfo {author} {\bibfnamefont {A.}~\bibnamefont
  {Okopinska}},\ }\href {\doibase 10.1103/PhysRevD.35.1835} {\bibfield
  {journal} {\bibinfo  {journal} {Phys. Rev. D}\ }\textbf {\bibinfo {volume}
  {35}},\ \bibinfo {pages} {1835} (\bibinfo {year} {1987})}\BibitemShut
  {NoStop}%
\bibitem [{\citenamefont {Duncan}\ and\ \citenamefont
  {Moshe}(1988)}]{Duncan:1988hw}%
  \BibitemOpen
  \bibfield  {author} {\bibinfo {author} {\bibfnamefont {A.}~\bibnamefont
  {Duncan}}\ and\ \bibinfo {author} {\bibfnamefont {M.}~\bibnamefont {Moshe}},\
  }\href {\doibase 10.1016/0370-2693(88)91447-5} {\bibfield  {journal}
  {\bibinfo  {journal} {Phys. Lett. B}\ }\textbf {\bibinfo {volume} {215}},\
  \bibinfo {pages} {352} (\bibinfo {year} {1988})}\BibitemShut {NoStop}%
\bibitem [{\citenamefont
  {Romatschke}(2019{\natexlab{c}})}]{Romatschke:2019rjk}%
  \BibitemOpen
  \bibfield  {author} {\bibinfo {author} {\bibfnamefont {P.}~\bibnamefont
  {Romatschke}},\ }\href {\doibase 10.1007/JHEP03(2019)149} {\bibfield
  {journal} {\bibinfo  {journal} {JHEP}\ }\textbf {\bibinfo {volume} {03}},\
  \bibinfo {pages} {149} (\bibinfo {year} {2019}{\natexlab{c}})},\ \Eprint
  {http://arxiv.org/abs/1901.05483} {arXiv:1901.05483 [hep-th]} \BibitemShut
  {NoStop}%
\end{thebibliography}%

\end{document}